# On the relation between the (censored) shifted Wald and the Wiener distribution as measurement models for choice response times

Robert Miller*[1,2], Stefan Scherbaum[1], Daniel W. Heck[3], Thomas Goschke[1], Sören Enge[4]

[1]Institute of Psychology, Technische Universität Dresden, Germany

[2]Department of Medical Epidemiology and Biostatistics, Karolinska Institutet, Stockholm, Sweden

[3]Department of Psychology, University of Mannheim, Germany

[4]Faculty of Natural Sciences, Medical School Berlin, Germany

*Correspondence concerning this article should be addressed to:

Dr. Robert Miller, TU Dresden, Fachrichtung Psychologie, Zellescher Weg 19, 01069 Dresden, Germany, Phone: +49 351 463 36835, Fax: +49 351 463 37274, Email: robert.miller@tu-dresden.de / robert.miller@ki.se

**Abstract**

Inferring processes or constructs from performance data is a major hallmark of cognitive psychometrics. Particularly diffusion modeling of response times (RTs) from correct and erroneous responses using the Wiener distribution has become a popular measurement tool because it provides a set of psychologically interpretable parameters. However, an important precondition to identify all of these parameters, is a sufficient number of RTs from erroneous responses.

In the present article, we show by simulation that the parameters of the Wiener distribution can be recovered even from tasks yielding very high or even perfect response accuracies using the shifted Wald distribution. Specifically, we argue that error RTs can be modeled as correct RTs that have undergone censoring by using techniques from parametric survival analysis.

We illustrate our reasoning by fitting the Wiener and (censored) shifted Wald distribution to RTs from 6 participants who completed a Go/No-go task. In accordance with our simulations, diffusion modeling using the Wiener and the shifted Wald distribution yielded identical parameter estimates when the number of erroneous responses was predicted to be low. Moreover, the modeling of error RTs as censored correct RTs substantially improved the recovery of these diffusion parameters when premature trial timeout was introduced to increase the number of omission errors. Thus, the censored shifted Wald distribution provides a suitable means for diffusion modeling in situations when the Wiener distribution cannot be fitted without parametric constraints.

1. Introduction

A fundamental aim of cognitive psychometrics relates to the development of measurement models that provide interpretable parameters from behavioral performance data. Technically, such measurement models are used to estimate the parameter manifestations for each person (and/or item) that maximize the likelihood of the probability distribution of their respective performance data. The functional form of this distribution is determined by the stochastic process that is assumed to generate the data. For example, in classical pen-and-paper tasks that consist of two-alternative forced choice (2AFC) items, a Bernoulli process is often thought to generate response accuracies (i.e., portions of correct responses across all items) that consequently follow a person-specific Binomial distribution (Lord, 1953). The parameter estimate(s) from this measurement model are interpretable in terms of the trait ability to respond correctly given the validity of its underlying process model.

The modern era of computerized tests, however, provided the opportunity to augment the analysis of response accuracies by accurately measured response times (RTs) and thus promoted the development of novel measurement models to integrate both kinds of performance data. To this end, the diffusion model (Stone 1960, Laming, 1968, Link & Heath, 1975, Ratcliff, 1978) and its derivatives have become increasingly popular for the simultaneous analysis of response accuracy and RT data from 2AFC tasks (for review see Voss, Nagler & Lerche, 2013). At its heart, the process component of the diffusion model assumes that the evidence accumulation for any response can be represented as a volatile drift towards either of two different boundaries. These boundaries characterize voluntarily adjustable decision thresholds whose transgression by the evidence-accumulation process initiates the execution of the respective correct or erroneous response (see section 2.1., or Voss et al., 2013, and Wagenmakers, 2009, for more detailed information).

The resulting joint distribution of responses and RTs (i.e., the corresponding measurement model; Vandekerckhove, Tuerlinckx, & Lee, 2011) is called the Wiener distribution, which depends on four psychologically meaningful parameters. These parameters are ($\alpha$) the distance between the two decision boundaries representing response caution, ($\beta$) the initial relative response tendency towards the correct response representing response bias, ($\delta$) the drift rate towards the correct response representing discriminability of both response options, and ($\theta$) the non-decision time, that is, the time that is not related to the evidence-accumulation but to sensory, motor, and other residual processes. Moreover, extensions of the basic diffusion model assume that $\beta$, $\delta$, and $\theta$ can vary across the items / trials of a 2AFC task (Ratcliff & Rouder, 1998; Ratcliff & Tuerlinckx, 2002). Thus the actual RT distribution of a person is assumed to be a mixture of many different Wiener distributions (Vandekerckhove et al., 2011).

In fact, there is good evidence showing that the Wiener distribution fits empirical performance data well, but all of its parameters are only empirically identified (i.e., uniquely estimable) if a sufficient number of RTs for both response options is available (Wagenmakers, van der Maas, & Grasman, 2007). Unfortunately, this precondition is either barely met in tasks that yield extremely low error rates (e.g., context interference tasks like Stroop or Simon tasks; Pratte, Rouder, Morey, & Feng, 2010), or even impossible to be met in tasks that yield no RTs for erroneous responses at all (e.g., response inhibition tasks like the Go/No-go task; Verbruggen & Logan, 2008).

In order to obtain interpretable parameter estimates in such situations, two rather suboptimal strategies have been used so far. First one can force the Wiener distribution to fit the data by constraining the starting point of the diffusion process (i.e. the response tendency) according to some a priori expectation (mostly $\beta = \alpha/2$ for ambiguous response tendencies; see Wagenmakers et al., 2007). While this approach is probably valid from a process modeling perspective, it is hardly suitable from a measurement perspective. This is because any constraining of β will result in a violation of the specific objectivity of parameter comparisons between individuals, or conditions within individuals that differ in their response tendencies (see Table 1 for illustration). The estimation of β by the simultaneously fitting of RTs from items / trials that probably share α as a common (person) parameter, but differ in their other parameters (e.g. trial types with varying difficulty) could in principle resolve this problem. However, this approach becomes impractical when there are only RTs from one trial type per individual.

--- insert Table 1 here ---

Second, one can switch to other measurement models and separately analyze response accuracy and RTs from the correct responses. As the characterization of correct RTs by measures of central tendency entails a substantial loss of information, the utilization of complex distributions (e.g., the exponentially-modified Gaussian) has strongly been encouraged to describe RTs (Balota & Yap, 2011; see Van Zandt, 2000, for an overview about different distribution candidates). However, Matzke and Wagenmakers (2009) found that the parameters of such distributions are not in strict conformity with the parameters of the Wiener distribution, thereby complicating specific interpretations in terms of an underlying diffusion process. This finding even generalized to the shifted Wald (also known as shifted inverse Gaussian) distribution that is generated by a diffusion process towards only one decision boundary as compared to the Wiener distribution that arises from two decision boundaries (Anders, Alario, & Van Maanen, 2016). Accordingly, the shifted Wald distribution is an acknowledged RT model for tasks with only one response option, but is also thought to have limited utility in 2AFC tasks for which the Wiener distribution has been the preferred measurement model[1].

Proceeding from these shortcomings, the present article investigates the diffusion modeling of RTs from correct *and* erroneous responses using the shifted Wald distribution. First we will show through simulation, that the parameters of the shifted Wald distribution cannot only approximate the parameters of the Wiener distribution whenever the response accuracy is high (as in the above mentioned tasks), but also when the RTs of the erroneous responses are incorporated as censored data into the shifted Wald distribution of correct RTs. Second, we will discuss this observation with regard to the interpretability of shifted Wald parameters in 2AFC tasks that promote high error rates (i.e., feature items / trials of high difficulty) by a narrowing of RT windows. The utility of (censored) shifted Wald modeling in such situations will be demonstrated with empirical data that has been obtained from a Go/No-go task.

**2. Theory: Correspondence of diffusion processes with one or two decision boundaries**

To increase the clarity of the following sections, we want to illustrate our reasoning by means of a Go/No-go task, which is commonly employed to assess the individual ability to inhibit habitual responses (Verbruggen & Logan, 2008). Response inhibition in Go/No-go tasks is challenged by the presentation of a randomly ordered trial sequence during which many Go-signals (e.g., consonants), and few No-go-signals (e.g., vowels) are consecutively presented on a computer screen. Typically, each of these trials requires participants to press a defined button in response to a Go-signal, but to omit this button press whenever a No-go-signal is encountered. More precisely, the response to No-go-signals is defined as an omission of the button press until an a priori defined time of trial timeout has elapsed (e.g., 1.5 seconds). Thus the response accuracy in both, Go- and No-go trials can be assessed, whereas only the RTs to correctly detected Go-signals can be analyzed.

---

[1] The two-boundary conceptualization of the standard diffusion model implies that the Wiener distribution of correct RTs transitions into a shifted Wald distribution if the contribution of the lower decision boundary (i.e., the number of erroneous responses) is negligible (see also section 2.1. and appendix). Only in such a case, the bivariate Wiener distribution for responses and RTs reduces to the univariate shifted Wald distribution.

Due to the higher frequency of Go-signals, most participants establish a habitual tendency to press the button as indicated by (1) a high number of erroneous responses to No-go-signals (i.e., No-go-signals are mistaken for Go-signals) and (2) comparably fast Go-signal RTs. The latter response acceleration is promoted by the task-inherent trial timeout, which would otherwise cause a considerable number of omitted button presses in Go-trials. Empirically, however, such errors can occur nonetheless for the two following reasons: First, a response may be *involuntarily* omitted due the accumulation of insufficient evidence for a confident button press in Go-trials prior to trial timeout. Second, Gomez, Ratcliff, and Perea (2007) argued that Go-trials may also yield errors because participants actually make the *voluntary* decision to omit the button press at a certain point prior to trial timeout (i.e., Go-signals are mistaken for No-go-signals). From a conceptual point of view, such a voluntary response forms no omission error, but a commission error that implies the erroneous detection of a No-go signal. However, the corresponding error RTs remain unobservable because only the subsequent trial timeouts are recorded.

In accordance with these two different error concepts, Go/No-go-tasks do not necessarily require only one response, that is, the detection of the Go-signal, but could as well represent a special type of a 2AFC task, where the actual responses to No-go-signals are unobservable (e.g., Trueblood, Endres, Busemeyer & Finn, 2011). This process-ambiguity makes the Go/No-go task a practical example to discuss the correspondence between the Wiener and the shifted Wald distribution as measurement models of responses and RTs from tasks that comprise either two, or only one response option, respectively.

## 2.1. Diffusion processes and the Wiener / shifted Wald distribution

From a process perspective, the Wiener and the shifted Wald distribution are generated by a volatile diffusion process that hits a boundary at different points in time (see appendix). In psychometric settings, such boundaries are thought to indicate the amount of evidence that is required for the initiation of a corresponding response. Thus, the Wiener and the shifted Wald distribution are conceptually equated with distribution of RTs in a given task. Figure 1 visualizes how both distributions can arise from the same diffusion process that starts to accumulate evidence for a response at mean drift rate $\delta$ after the time $\theta$ has elapsed without any evidence accumulation.

Notably, the Wiener distribution requires the existence of two boundaries $\gamma_u > \gamma_l$, which represent two different response options. In the modern literature on diffusion modeling (e.g. Wagenmakers, 2009), these two boundaries are mostly parametrized in terms of the separation between the two boundaries $\alpha = \gamma_u - \gamma_l$, and the relative starting point of the diffusion process $\beta = -\gamma_u / (\gamma_u - \gamma_l) + 1$. Conditional on the validity of the process model, the four parameters of the Wiener distribution can then be interpreted as ($\alpha$) response caution, ($\beta$) response bias towards the correct response option, ($\delta$) the mean accumulation rate of evidence for the correct response option, and ($\theta$) the non-decision time during which no evidence is accumulated. The probability density at the lower boundary of an accordingly parameterized Wiener distribution of RTs is given by Eq. 1 (Navarro & Fuss, 2009; see also Gondan, Blurton & Kesselmeier, 2014). Using Eq. 1, the probability density of the Wiener distribution at the upper boundary can also be obtained by substituting $\delta' = -\delta$ and $\beta' = 1 - \beta$.

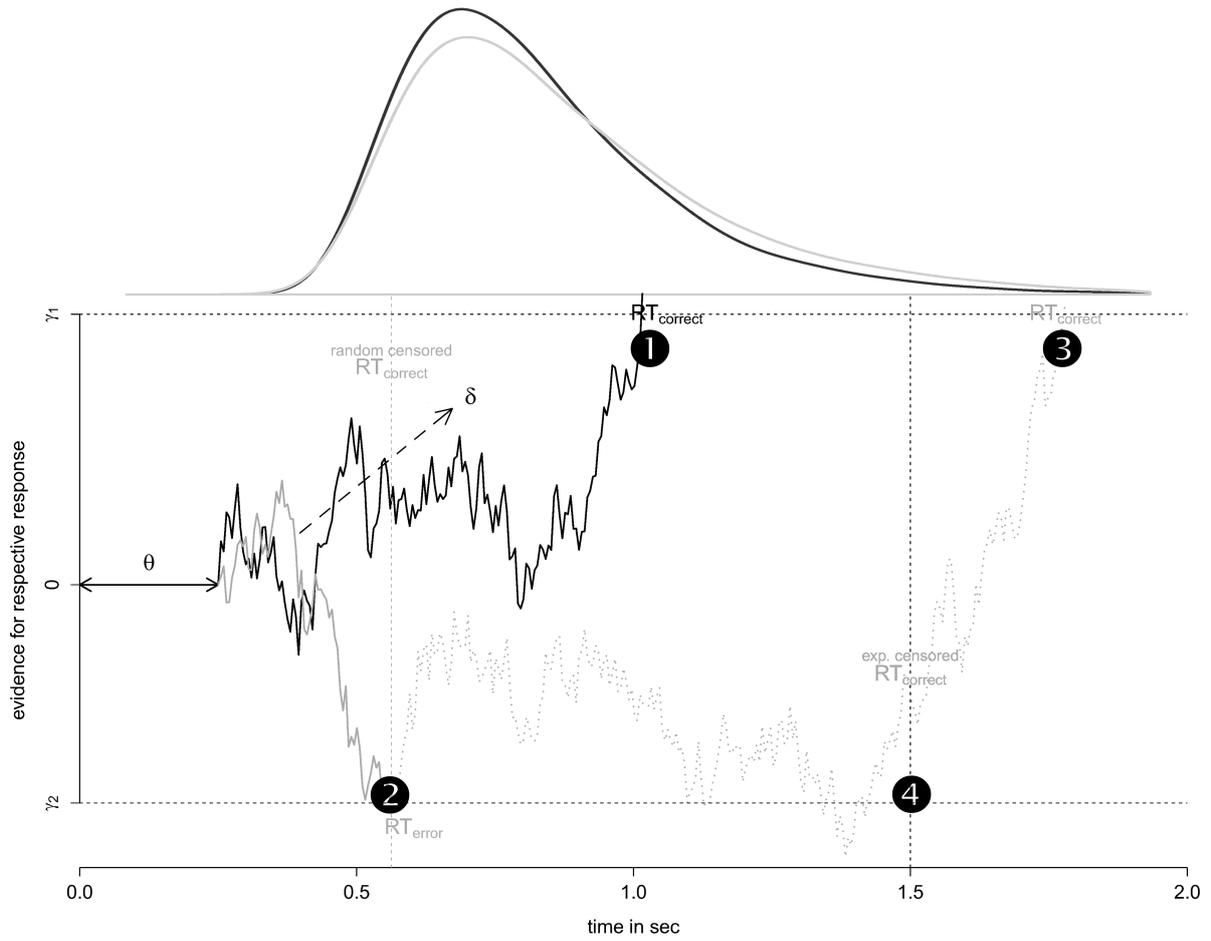

Figure 1. Illustration of a diffusion process that can generate both, a shifted Wald distribution (grey), and a Wiener distribution (black) of RTs. Note: If two boundaries exist ($\gamma_u$ and $\gamma_l$), the diffusion process will stop when it hits either boundary (solid paths) and generate a Wiener distribution comprising ❶ $RT_{correct}$ and ❷ $RT_{error}$. If only an upper boundary ($\gamma_u$) exists, the process will not stop when it hits $\gamma_l$, but continue until $\gamma_u$ will be finally reached (dotted grey path). This will then generate a shifted Wald distribution comprising ❶ $RT_{correct}$ and ❸ $RT_{correct}$. However, the process is often experimentally terminated at a certain point in time ❹, where ❸ is not observable but known to be located in between ❹ and ∞. Analogously, ❷ forms a randomly varying censoring threshold, where ❸ is known to be located in between ❷ and ∞.

Eq. (1)

$$f_W(t \mid \alpha, \beta, \delta, \theta) = \frac{\pi}{\alpha^2} \exp\left[-\alpha\beta\delta - \frac{1}{2}\delta^2(t-\theta)\right] \sum_{k=1}^{\infty} k \exp\left[-\frac{k^2\pi^2(t-\theta)}{2\alpha^2}\right] \sin(k\pi\beta)$$

By contrast, the shifted Wald distribution is generated if the process exclusively stops when it hits an upper boundary $\gamma_u$ representing the only available response option in a given task (i.e, $\gamma_l$ does not exist in this scenario). The resulting distribution of RTs is characterized by the probability density function given by Eq. 2 (Heathcote, 2004; Anders et al., 2016). Shifted Wald-distributed RTs have a more pronounced skew than implied by the Wiener distribution because a transgression of $\gamma_l$ cannot result in the premature termination of the diffusion process. Consequently, the portion of prolonged RTs at $\gamma_u$ increases as compared to the Wiener distribution.

Eq. (2)

$$f_{sW}(t \mid \gamma, \delta, \theta) = \frac{\gamma}{\sqrt{2\pi(t-\theta)}} \exp\left[-\frac{(\gamma - \delta t + \delta\theta)^2}{2(t-\theta)}\right]$$

A verbal illustration of the conditions under which both distributions approximate each other will henceforth be made using the above outlined Go/No-go task example: In any Go-trial, the diffusion process is supposed to drift towards the upper boundary $\gamma_u$ and stop when sufficient evidence for presence of the Go-signal has been accumulated, that is, the button press will be initiated when the process hits $\gamma_u$ (see solid black path in Figure 1). In the 2AFC scenario advocated by Gomez et al. (2009), the lower boundary $\gamma_l$ for *voluntarily* deciding to omit the button press (i.e., a commission error; the Go-signal is mistaken for a No-go signal) can only be hit due to the volatility of the process (see solid grey path in Figure 1). Given these two boundaries, the RT distribution of correct responses and of (unobservable) commission errors in Go-trials corresponds to the Wiener distribution. Hence all error-yielding manifestations of the diffusion process cannot contribute to forming the RT distribution of correct responses to Go-signals.

In the scenario with only one response option, by contrast, the same path of accumulated evidence could in theory also arise, but evidence accumulation would *not* stop when the lower boundary ($\gamma_l$) was reached (see vertical grey line in Figure 1). Instead, the diffusion process would further continue to drift towards the upper boundary ($\gamma_u$), implying the inevitable detection of any Go-signal at some delayed point in time (see dotted grey path in Figure 1) by means of a button press. Thus, errors will exclusively arise in this scenario by the *involuntary* omission of the button press, if the process fails to reach $\gamma_u$ before the experimentally controlled point of trial timeout (i.e., an omission error; see vertical black line in Figure 1). Under such circumstances, however, the shifted Wald distribution cannot completely account for the correct Go-signal RTs anymore unless all according omission errors are modeled as correct RTs that have become unobservable (i.e., censored) at the point of trial timeout (see Ulrich & Miller, 1991).[2]

Proceeding from this reasoning a first conclusion can be drawn: The density of the Wiener distribution at the upper decision boundary and the density of the shifted Wald distribution are supposed to be identical whenever neither omission errors, nor commission errors change the distribution of correct RTs (see Anders et al., 2016). A mathematical justification for this claim is provided in the appendix of this article.

## 2.2. The censoring of correct RTs by omission and commission errors

Intriguingly, the major conceptual analogy between the omission errors postulated by the shifted Wald distribution and the commission errors postulated by the Wiener distribution, is that any error type ultimately determines the point in time at which the correct RTs have become unobservable (i.e., the censoring point). Thus, any error only indicates that a participant was neither aware of the correct response at trial timeout, nor at an earlier (but unobservable) time of error commission. Nonetheless, in both cases the correct response would have been initiated later on (with a probability given by the shifted Wald distribution), if the diffusion process had neither been stopped by a timeout, nor by a commission error. This suggests a second conclusion: Whenever errors are present, their modeling as censored correct RTs using the shifted Wald distribution should yield parameter estimates that correspond to those obtained using the Wiener distribution. Thus, the shifted Wald distribution could be used for approximating the psychologically meaningful parameters of the Wiener distribution.

---

[2] In fact, the Wiener model can also not account for response omission without further modifications. For reasons of clarity, however, we will assume for now that the timeout occurs sufficiently late to enable the complete initiation of correct and error responses.

The term "censoring" has been popularized in the field of survival analysis and denotes missing data problems, in which only the interval of the unobserved time-to-event data is known (Miller, 1998). Such situations arise whenever the event of interest (e.g., the response to a Go-signal) will eventually have occurred outside the period of data collection (e.g., after trial timeout). In order to consider such censoring information appropriately, the probability density of observed correct RTs needs to be multiplied with the probability that correct RTs occur after trial timeout and are thus unobserved. According to the diffusion model, this probability is given by the survival function of the shifted Wald-distributed correct RTs, which is provided in Eq. 3 (see Heathcote, 2004).

Eq. (3)

$$S_{sW}(t \mid \gamma, \delta, \theta) = 1 - F_{sW}(t \mid \gamma, \delta, \theta) = 1 - \Phi\left(\frac{\delta t - \delta\theta - \gamma}{\sqrt{t - \theta}}\right) - e^{2\gamma\delta}\Phi\left(-\frac{\delta t - \delta\theta + \gamma}{\sqrt{t - \theta}}\right)$$

Here, $\Phi$ denotes the distribution function of the unit normal distribution. For all $n$ observable correct RTs and censored correct RTs (i.e. the point of trial timeout), the likelihood function is then given by Eq. 4 (Miller, 1998; see also Ulrich & Miller, 1994), where $d$ is a censoring indicator with $d = 1$ for all correct RTs and $d = 0$ for all censored correct RTs.

Eq. (4)

$$L(\gamma, \delta, \theta \mid t) = \prod_{i=1}^{n} f_{sW}(t_i|\gamma, \delta, \theta)^d \prod_{i=1}^{n} S_{sW}(t_i|\gamma, \delta, \theta)^{1-d}$$

This likelihood of the censored shifted Wald distribution is identical to the likelihood of the conventional shifted Wald distribution of correct RTs whenever there are no censored correct RTs. Crucially, however, the correspondence of the parameter estimates that maximize this likelihood function given the RT data, is restricted to the case of independent mechanisms generating the observable and censored correct RTs (i.e. censoring needs to be non-informative; e.g. Siannis, Copas & Lu, 2005). While this assumption would be reasonable if only one boundary existed, the standard diffusion model postulates that any evidence accumulation in favor of the correct response option is concurrently evidence against the erroneous response option[3]. Thus, the evidence accumulation for correct responses, and unobservable commission errors needs to be performed in two separate cognitive instances. Otherwise the diffusion parameter estimates will be biased as a function of the proportion of censored observations.

---

[3] Several other models (e.g. the race-model for stop-signal tasks, Logan et al., 2014, or the linear ballistic accumulator, Brown & Heathcote, 2008) disagree with the response competition implied by the standard diffusion model and also assume independent evidence accumulation processes.

This resembles to a competing risks situation (Prentice et al., 1978) in which the (sub)distributions of the correct RTs and the error RTs depend on the same parameters. Specifically, the non-decision time θ is fully shared by both distributions, whereas a drift $\delta_1$ towards the upper boundary $\gamma_u$ is concurrently a drift away ($\delta_2 = -\delta_1$) from the lower boundary $\gamma_l$. Consequently, the joint likelihood function of correct RTs and error RTs is the product of the two separate likelihood functions of both RT distributions. This function is provided in Eq. 5, in which all RTs from the correct response option are used as censoring points for the incorrect response option and vice versa (see Prentice et al., 1978; Miller, 1998).

Eq. (5)

$$L(\gamma_u, \gamma_l, \delta, \theta \mid t)$$

$$= \underbrace{\prod_{i=1}^{n} f_{sW}(t_i \mid \gamma_u, \delta, \theta)^d \prod_{i=1}^{n} S_{sW}(t_i \mid \gamma_u, \delta, \theta)^{1-d}}_{likelihood\ of\ correct\ RTs} \underbrace{\prod_{i=1}^{n} f_{sW}(t_i \mid \gamma_l, -\delta, \theta)^{1-d} \prod_{i=1}^{n} S_{sW}(t_i \mid \gamma_l, -\delta, \theta)^d}_{likelihood\ of\ error\ RTs}$$

In the following section, we show through simulation that this censored shifted Wald model can recover diffusion parameters based on RTs that are actually sampled from the Wiener distribution (i.e., they are generated by a diffusion process with two boundaries).

### 3. Simulation: Recovery of Wiener parameters by (censored) shifted Wald modeling

Proceeding from the outlined correspondence of diffusion processes with one or two boundaries, we performed a simulation study to demonstrate how the parameters obtained by both, conventional and censored shifted Wald modeling relate to different parameter configurations of the Wiener distribution. All shifted Wald parameters and their corresponding Wiener parameter configurations are listed in Table 2.

--- insert Table 2 here ---

In a first step, we generated artificial RTs from a large set of Wiener parameters covering the range of previously reported estimates (Matzke & Wagenmakers, 2009). Across the parameter ranges reported by Matzke and Wagenmakers (2009) we calculated 1,000 equidistant manifestations of each parameter of the Wiener distribution (see Table 2, e.g. $\theta_1 = 0.21$ s, $\theta_2 = 0.21073$ s, $\theta_3 = 0.21146$ s, …, $\theta_{1000} = 0.94$ s). These parameter manifestations were appended to the respective difference set of reference parameter manifestations ($\alpha = 2$, $\beta = 0.5$, $\delta = 2.5$ s$^{-1}$, $\theta = 0.55$ s), which resulted in 4 x 1,000 different diffusion scenarios. We then randomly generated 1,000 artificial RTs comprising both, correct and erroneous responses, for each of these scenarios.

In a second step, these data were submitted to three different types of shifted Wald modeling. Specifically, the parameters maximizing the likelihood function (MLE) of the conventional shifted Wald distribution, the censored shifted Wald distribution (Eq. 4), and the shifted Wald distribution for competing risks (Eq. 5) were estimated using the SIMPLEX algorithm (Nelder & Mead, 1968) with starting values that have been estimated using the method of moments (Heathcote, 2004).

Finally, these estimates were regressed on the known parameters of the data-generating Wiener distributions using locally weighted scatterplot smoothing (Cleveland & Devlin, 1988). All simulations were conducted using the RWiener package (Wabersich & Vandekerckhove, 2014) with R 3.2.2 statistical software (R Core Team, 2015).

### 3.1. Trial-invariant diffusion parameters

First, we investigated the case where all randomness in the RT distributions is attributable to the inherent volatility of the diffusion process, implying Wiener distributions with no parameter variability across trials. These simulations' results are presented in Figure 2.

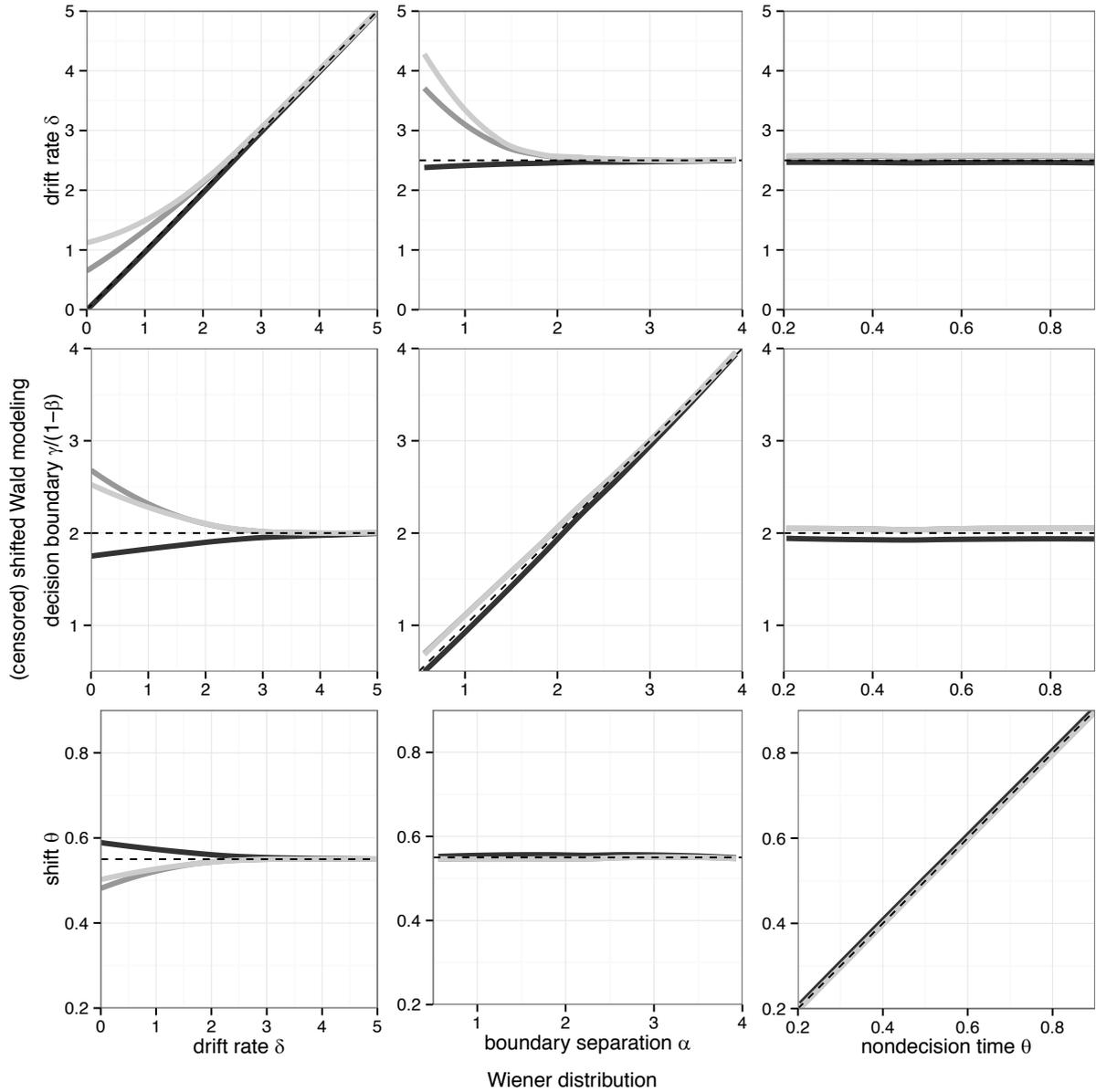

Figure 2. Recovery of Wiener parameters by different types of shifted Wald modeling. Note: In all simulations, one parameter of the Wiener distribution was systematically varied (see section 3). Lines show parameter recovery by conventional shifted Wald modeling (light grey), censored shifted Wald modeling (dark grey), and censored shifted Wald modeling with competing risks (black).

Regarding the parameter recovery by conventional shifted Wald modeling (light grey lines), the simulations revealed pronounced linear relations between the corresponding shifted Wald and Wiener parameters when the proportion of correct responses amounted to more than ~95%[4]. However, our results also confirmed a growing divergence of Wiener and shifted Wald parameters as the proportion of correct responses decreased due to narrow boundary separations ($\alpha$) and/or small drift rates ($\delta$). Specifically, the latter also led to biased estimates of boundary separations and non-decision times ($\theta$). Likewise, response tendencies ($\beta$) toward the erroneous response option entailed a growing discrepancy between parameter estimates (data not shown).

In accordance with our reasoning, censored shifted Wald modeling (dark grey lines) was well able to improve the recovery of $\alpha$ and $\delta$, but the bias of $\alpha$ and $\theta$ resulting from a small drift of the data-generating diffusion process could not be adjusted for. The failure of censored shifted Wald modeling to fully recover Wiener parameters at high error rates is not surprising, because any censoring induced by erroneous responses will be informative for a delayed execution of the correct responses (see Figure 1) unless the processes governing correct and erroneous responses were independent (cf. Jones & Dzhafarov, 2014). Such informative censoring is well known to introduce bias when the likelihood function of non-informatively censored distributions is maximized (e.g., Siannis et al., 2005).

In order to alleviate this issue, we implemented censored shifted Wald modeling that considered the probability of error commission as a competing risk for correct responses.

---

[4] Notably, Matzke and Wagenmakers (2009) were not able to discover these linear relations between the shifted Wald and Wiener parameters because they relied on an absolute parameterization of the starting point of the diffusion process that was more common at that time.

Thus, we modeled both the joint distribution of correct RTs and error RTs using two censored shifted Wald distributions. Both RT distributions feature different boundaries ($\gamma_u$ and $\gamma_l$, respectively), which depend on a shared response tendency parameter and the distance in between them (see Table 2). Moreover, the distribution of error RTs was constrained to arise from the negative drift parameter of the distribution of correct RTs (i.e. $\delta_1 = \delta$ and $\delta_2 = -\delta$). Thus, these two distributions mimicked the Wiener model as two connected evidence accumulators. As can be seen in Figure 2 (black lines), the competing risks approach resulted in an almost perfect recovery of the Wiener parameters when error rates were not negligible, with small bias only occurring when estimating $\alpha$ at extremely small drift rates $\delta$. By contrast, the recovery of all Wiener parameters was insensitive to manipulations of $\alpha$ and $\theta$. However, we need to emphasize, that the increased accuracy of the competing risks approach at higher error rates came at the cost of the same limitations as the Wiener model of 2AFC tasks, that is, any response tendency $\beta$ that is not identified by the data will necessarily entail an unidentifiable boundary separation $\alpha$ (or boundaries $\gamma_u/\gamma_l$) under conditions of low error rates.

**3.2. Diffusion parameter variability across trials**

RT modeling from only one response option (e.g. the correct response) does not necessarily require variable diffusion parameters because RT variability can just arise from the stochasticity in the process itself (Jones & Dzhafarov, 2014). With two competing response options, however, the Wiener distribution can only account for empirical phenomena if parameter variability across different trials is explicitly assumed (Ratcliff & Rouder, 1998). In consequence we complement the simulations in section 3.1 by investigating the sensitivity of (censored) shifted Wald modeling of performance data to such parameter variability. The results of these simulations are presented in Figure 3 and essentially replicate the findings of Matzke and Wagenmakers (2009): The solid lines show that conventional and censored shifted Wald analyses are comparably robust to large variability in drift rates and to moderate variability in response tendencies across trials. However, particularly non-decision time variability is able to severely bias the recovery of the data-generating Wiener parameters. The reason for these findings relates to the strict assumption of the shifted Wald likelihood function that there cannot be any RTs smaller than the non-decision time parameter θ. Thus, the minimum RT constrains the upper bound of θ irrespective of the possibility that these fast RTs are probably only contaminants that obscure the actual data-generating process. In order to provide a means to account for variability in non-decision time, the censored shifted Wald model can be extended by directly modeling the additive, random non-decision time θ. Specifically, if the random variable $T_1$ follows a (censored) non-shifted Wald distribution and the independent non-decision time $T_2$ follows a uniform distribution on the range $[\theta_{min}, \theta_{max}]$, the density of the observed RT random variable $T = T_1 + T_2$ is given by the convolution:

Eq. (6)

$$f_{sW\theta}(t \mid \gamma, \delta, \theta_{min}, \theta_{max}) = \int_0^t f_W(\theta \mid \gamma, \delta) u(t - \theta \mid \theta_{min}, \theta_{max}) d\theta$$

$$= \frac{1}{\theta_{max} - \theta_{min}} \int_{t-\theta_{max}}^{t-\theta_{min}} f_W(\theta \mid \gamma, \delta) \, d\theta = \frac{F_{sW}(t|\gamma, \delta, \theta_{min}) - F_{sW}(t|\gamma, \delta, \theta_{max})}{\theta_{max} - \theta_{min}}$$

Note that this density can also be interpreted as the integral over a continuous uniform mixture of shifted Wald distributions (Van Zandt & Ratcliff, 1995). The dashed red lines in Figure 3 illustrate that any bias in the parameter recovery (resulting from variability in non-decision times) can be removed when the censored version of this extended shifted Wald model is used.

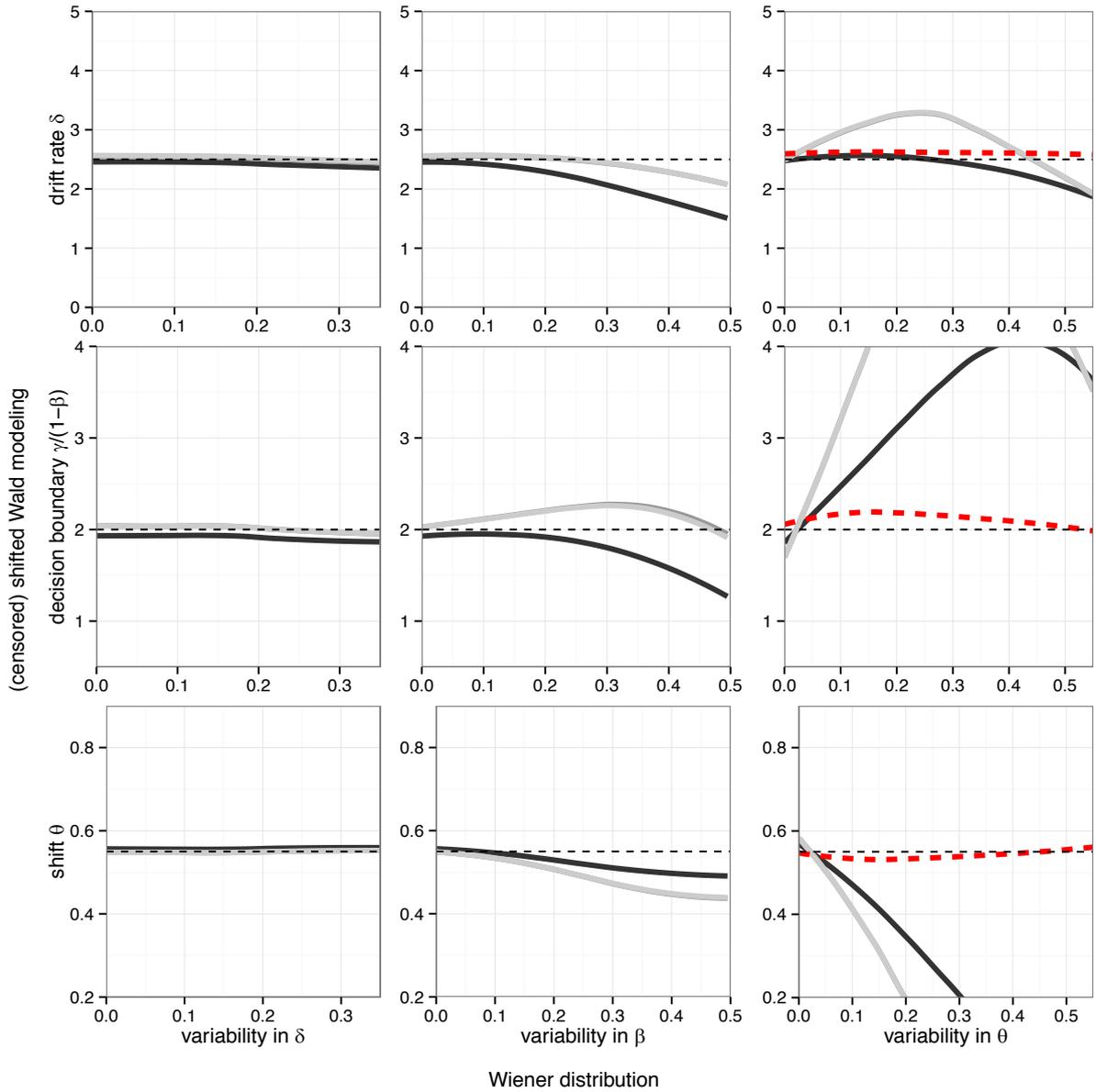

Figure 3. Sensitivity of the shifted Wald modeling to potential variability in Wiener parameters across trials *i*. Note: Proceeding from the reference manifestations at $\delta = 2.5$ s$^{-1}$, $\beta = 0.5$, and $\theta = 0.55$ s (horizontally oriented, dashed lines), parameter variability was systematically introduced according to the common distribution assumptions of the Wiener diffusion model (see Jones & Dzhafarov, 2014): $\delta_i \sim \text{Normal}(\delta, \sigma_\delta)$, $\beta_i \sim \text{Uniform}(\beta - \sigma_\beta, \beta - \sigma_\beta)$, and $\theta_i \sim \text{Uniform}(\theta - \sigma_\theta, \theta - \sigma_\theta)$. Lines show recovery by conventional/censored shifted Wald modeling (light/dark grey), censored shifted Wald modeling with competing risks (black), and censored shifted Wald modeling with uniform additive non-decision times (red dashed).

Besides the variability in non-decision time that should be modeled appropriately (Ratcliff & Turlinckx, 2002; Wagenmakers et al., 2008), fast RTs are particularly plausible in empirical settings because of prepotent trials (e.g. Go-trials). Crucially, such trials promote response bias that may be carried over to other trial types. Logan, Van Zandt, Verbruggen, and Wagenmakers (2014) derived a version of the Wald density function that can account for fast RTs by assuming an upper decision boundary γ that varies uniformly across trials. A shifted Wald version of this density function is provided in Eq. (7), where $\phi$ and $\Phi$ and denote the density function and the distribution function of the unit normal distribution, respectively.

Eq. (7)
$$f_{sW\gamma}(t \mid \gamma_{min}, \gamma_{max}, \delta, \theta)$$
$$= \frac{1}{\gamma_{max} - \gamma_{min}} \left[ \phi\left(\frac{\delta t - \delta\theta - \gamma_{min}}{\sqrt{t - \theta}}\right) - \phi\left(\frac{-\delta t + \delta\theta + \gamma_{max}}{\sqrt{t - \theta}}\right) \right.$$
$$\left. - \delta\left(\Phi\left(\frac{\delta t - \delta\theta - \gamma_{min}}{\sqrt{t - \theta}}\right) - \Phi\left(\frac{-\delta t + \delta\theta + \gamma_{max}}{\sqrt{t - \theta}}\right)\right) \right]$$

Irrespective of these viable solutions, we like to point out that the modeling of parameter variability primarily serves to filter the central tendency of the respective diffusion parameters in the presence of between-trial noise, but can itself hardly be recovered from a reasonable number of trials (Van Ravenzwaaij & Oberauer, 2008).

**4. Example: Shifted Wald and Wiener modeling of performance in a Go/No-go task**

So far, we have argued that the shifted Wald and Wiener parameters are actually comparable (A) in the presence of high drift rates, high decision boundaries γ, or response tendencies toward the correct response option[5], or (B) if the competition between correct and erroneous responses is adequately modeled.

In situation (A), the number of errors will be very low. Therefore, conventional and censored shifted Wald analyses will asymptotically yield the same parameter estimates, and one will probably trade off the additional implementation burden of fitting the censored shifted Wald distribution against the prospected gain. Conversely, whenever a task yields a moderate number of commission errors (i.e. > 5% that do not result from trial timeout), the number of error RTs will suffice to estimate reasonable diffusion parameters using both, the Wiener model, and the (competing risks variant of the) censored shifted Wald model. Thus, the competing risks approach to censored shifted Wald modeling is mostly interesting from a theoretical perspective, but often of less practical value for dealing with situation (B).

However, the modeling of error RTs as uninformatively censored correct RTs may have immense advantages if there are a considerable number of errors due to involuntary response omission. This situation is often caused by the fast trial timeouts in typical Go/No-go tasks, which is a common means to experimentally increase the task difficulty. Proceeding from this situation, we illustrate in this section that censored shifted Wald modeling can actually serve to estimate diffusion parameters more accurately as compared to fitting the Wiener distribution.

**4.1. Material and methods**

In order to accomplish this goal, we present the results from (conventional and censored) shifted Wald modeling and Wiener modeling of 350 valid Go-trial RTs provided by each of six undergraduate students, who completed a Go/No-go task that was modified for providing both, correct and commission error RTs. Each trial of this task required to respond by pressing the key "Y" to any presented Go-signal (two horizontally aligned circles), and to press the key "M" whenever the No-go-signal (two vertically aligned circles) was presented. The proportion of Go-trials amounted to 87.5%, yielding a total number of 400 Go- and No-go trials. Each trial was subjected to a timeout at 1.25 seconds, which was not exceeded by any participant. Thus, no omission errors were observed. Moreover, all Go-trial response accuracies amounted to 100%, implying that no errors were committed because of mistaking of a Go-signal for a No-go-signal at a certain point in time.

---

[5] Please note, that both situations also assume that across-trial variability in response tendencies or non-decision times is negligible. If this assumption is unlikely to hold, such variability should be accounted for appropriately.

While shifted Wald modeling is ideally suited to estimate diffusion parameters in such situations, the standard diffusion model is not identified. One common way to circumvent this issue requires a constraint of the response tendency parameter of the Wiener distribution ($\beta$) to a predetermined value. Although it seems intuitive to fix $\beta$ to the proportion of Go-trials (i.e., $\beta = .875$), this value would nonetheless be arbitrary guess. Thus, we evaluated the impact of several (more or less likely) response biases varying in between $0.1 \leq \beta \leq 0.9$. As explained in section 3, the boundary separation $\alpha$ of the Wiener distribution is supposed to relate to the decision boundary $\gamma$ of the shifted Wald distribution by $\gamma = \alpha(1-\beta)$ whenever there are no error RTs. Upon presence of pronounced response tendencies towards the Go-signal, the Wiener distribution is expected to transition into the Wald distribution if drift rates and boundary separations are sufficiently large (see Figure 2).

In order to simulate scenarios in which the number of omission errors would have been higher, we further evaluated the impact of post-hoc censoring (i.e., truncation) on the estimated diffusion parameter. Specifically, we truncated all correct RTs that exceeded (virtual) trial timeouts at 300 ms, 400 ms, and 500 ms, respectively, thereby generating artificial omission errors. The likelihood of the diffusion parameters that were estimated from these censored data using the conventional shifted Wald/Wiener model given the original RT data was thereafter contrasted to the likelihood of the parameters estimates from the censored shifted Wald model. In other words, we investigated if the explicit modeling of the censoring mechanism was able to reduce the bias of diffusion parameter estimates that results from the inability to account for actual omission errors using the Wiener distribution. Again, maximum likelihood parameter estimates (MLE) were obtained for all model variants using the SIMPLEX algorithm (Nelder & Mead, 1968), the RWiener package (Wabersich & Vandekerckhove, 2014) and R 3.2.2 statistical software (R Core Team, 2015). The syntax and the data are provided as supplementary material to this article.

### 4.2. Results

The original performance data and the according diffusion parameter estimates are listed in the upper part of Table 3. For all six participants, the fit of the shifted Wald model is visualized in Figure 4A. Consistent with our prediction, we found that the diffusion parameters estimated by conventional shifted Wald and Wiener modeling from the original RT data corresponded perfectly for all participants, with means (± standard deviations) of $\delta$ = 6.74 (±1.70) s$^{-1}$, $\theta$ = 0.17 (±0.03) s, and $\gamma$ = 1.13 (±0.19) corresponding to $\alpha$ = 2.26 (when $\beta$ = 0.5) or $\alpha$ = 5.65 (when $\beta$ = 0.8). The only prerequisite for this approximation was a response bias towards the Go-signal (i.e. $\beta \geq 0.5$; see Figures 4B/4C). The censored shifted Wald approach yielded the same estimates as the shifted Wald and the Wiener model, because all Go-stimuli were correctly identified (i.e. there were no censored correct RTs).

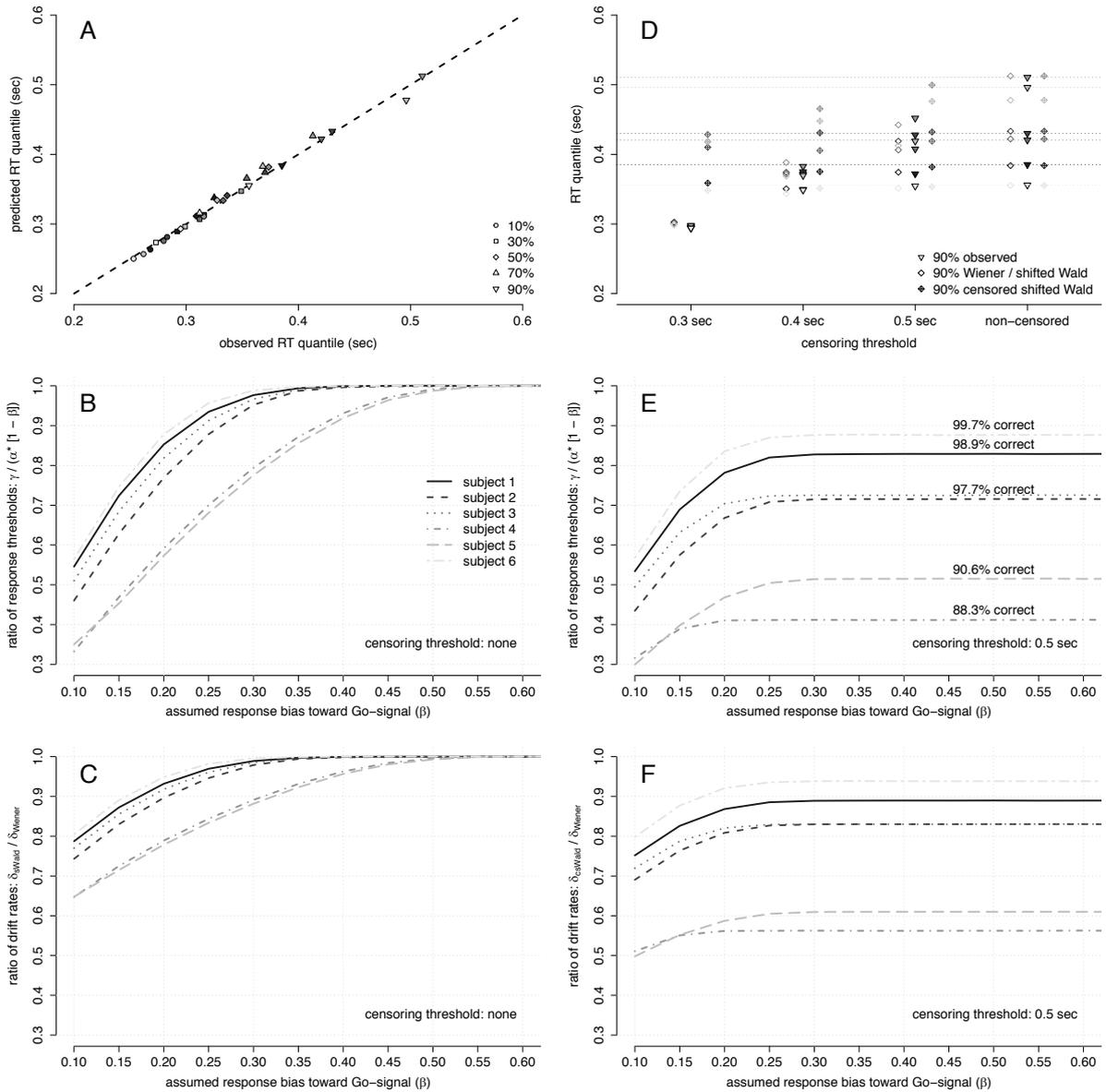

Figure 4. Performance of Wiener vs. censored shifted Wald modeling under conditions of timeout-related response omission. Note: Panel A shows the observed RT quantiles of 350 Go-trials (for 6 subjects) that are plotted against the RT quantiles predicted by the shifted Wald distribution. Wiener modeling yields exactly the same parameter estimates (B/C) and thus identical predictions (D) as the shifted Wald model whenever there is a response bias toward reporting presence of the Go-Signal (i.e., β > 0.5). Panel D illustrates how the modeling of artificially generated omission error RTs yields improved predictions of the original RT quantiles when the censored shifted Wald distribution is fitted to these data. The bias of these predictions depends on the number of omission errors (E/F).

--- insert Table 3 here ---

After artificially censoring these data, we observed a discrepancy between the predictions of the conventional shifted Wald/Wiener model, and the censored shifted Wald model that increased as a function of decreasing trial timeout (see Figure 4D). Notably the latter yielded parameter estimates that were much closer to the "true" parameters (assuming that they were accurately estimated from the original, non-censored RT data) than those estimated by conventional shifted Wald/Wiener modeling of the remaining, correct RTs. This is also reflected by the likelihood ratios listed in the last column of Table 3, which highlight that the parameters of the censored shifted Wald model were at least twice as likely as those estimated by the conventional shifted Wald model given the original performance data. Accordingly, the parameter ratio between the censored shifted Wald and the conventional shifted Wald/Wiener approach decreased as a function of censored RTs (Figure 4E/F), implying that drift rates, response boundaries, and non-decision times were substantially underestimated whenever these censored RTs were not taken into account.

## 6. Discussion

In the present article, we showed through simulation that the parameter estimates from shifted Wald modeling approximate those of the Wiener distribution if the number of erroneous responses is low (i.e. < 5%) due to high drift rates, high decision boundaries, and a-priori response tendencies towards the correct response option. Considering that these conditions seem to be met in many cognitive tasks that actually involve two response options (e.g., Gomez et al., 2007; Pratte et al., 2010), shifted Wald modeling often provides interpretable diffusion parameters, while bypassing the need to impose more or less arbitrary constraints on unidentifiable parameters when using the Wiener distribution. Notably, the approximation of the Wiener by the shifted Wald distribution given low error rates arises because both measurement models assume the same diffusion process of sequential evidence accumulation. By contrast, other distributions can hardly be used to estimate parameters that are analogously interpretable (see Matzke & Wagenmakers, 2009; Balota & Yap, 2011).

Moreover, we demonstrated that any restriction of RT windows in Go/No-go tasks would introduce a censoring mechanism that fosters biased (i.e. less interpretable) diffusion parameter estimates. Besides an allocation of sufficient time for correct stimulus processing (i.e., trials should not to be censored prior to task completion, Ulrich & Miller, 1994), we argue for the modeling of error RTs as censored correct RTs in order to deal with this issue. Importantly, the modeling of censored correct RTs can be easily implemented using the shifted Wald distribution. By contrast, the readily available implementations of the Wiener distribution do not consider this option, because they posit that erroneous responses are mostly generated by unobservable commission errors. However, such commission errors cannot be distinguished from actual omission errors in typical Go/No-go tasks. Interestingly, a conceptually analogous method has been proposed to model unobservable responses in Stop-signal tasks (Logan et al., 2014).

Despite of these advantages of (censored) shifted Wald modeling, our simulations also revealed that parameter estimates become biased if drift rates are small, decision boundaries are small, or if there is a strong response tendency towards the erroneous response option. Although all of these three sources of bias can potentially restrict the interpretability of shifted Wald parameters as diffusion parameters, the presented Go/No-go data suggest that bias is unlikely to occur when using an appropriate task design. Next to the prepotent response category that is an inherent feature of all response inhibition tasks, overly small decision boundaries can also be avoided in situations involving a balanced number of response options by instructing the participants to focus on response accuracy (as opposed to speed; cf. Wagenmakers, 2009). Irrespective of these considerations, we want to emphasize that the Wiener distribution will necessarily remain the measurement model of choice to infer on diffusion parameters when a sufficient number of error RTs is actually available.

Finally, we need to point out that a single trial-invariant shifted Wald distribution may not completely account for empirical phenomena in 2AFC tasks. This has been the major reason for introducing across-trial variability of diffusion parameters in the first place (Ratcliff & Rouder, 1998, Ratcliff & Turlinckx, 2002). In consequence, it is not surprising that shifted Wald modeling will probably fail to recover diffusion parameters if such variability is explicitly assumed[6] (Matzke & Wagenmakers, 2009). This was shown to particularly limit the applicability of (censored) shifted Wald modeling when non-decision times (and to a lesser extent response tendencies) vary substantially across trials. Apart from the modeling of non-decision times as a uniform additive random variable (as proposed in section 3.2.), one could also consider to model RTs as a mixture of the "true" shifted-Wald-distributed RTs and uniformly distributed response contaminants to resolve this issue (Ratcliff & Tuerlinckx, 2002; Wagenmakers et al., 2008). These approaches could be further extended by modeling a possible between-trial variability of decision boundaries (see Logan et al., 2014), highlighting that the (censored) shifted Wald distribution actually has the potential to serve as a generic diffusion measurement model for choice response times.

**Appendix**

In cognitive psychometrics, the term *diffusion* refers to a stochastic process that is thought to describe the change of accumulated evidence $x$ across the time $t$ (scaled in seconds) by the differential equation $dx / dt \sim \text{Normal}(\delta, 1)$. Here $p(x, t \mid \delta)$ shall denote the probability that a manifestation of this process is located at *x* after *t* has elapsed. Assuming that such a process starts at $x = 0$, $t = 0$, and stops when it hits a response boundary $\gamma_u > 0$, the probability that the process has not stopped before *t* is given by the survival probability of $p(x, t \mid \delta)$:

$$S(t \mid \delta, \gamma_u) = \int_{-\infty}^{\gamma_u} p(x, t \mid \delta) dx$$

---

[6] As Wagenmakers and colleagues (2008) aptly remarked "…it is easy to generate data from a complex model and show that the simpler model, nested within the complex model, fails to recover parameters well".

The density function of the Wald distribution can be derived from this survival probability as $f_{SW}(t \mid \gamma_u, \delta,) = dS(t \mid \delta, \gamma_u) / dt$.

In contrast to this one-boundary-scenario that generates the Wald distribution, the Wiener distribution is generated by a diffusion process that stops when it hits either of two boundaries $\gamma_u > 0 > \gamma_l$. The corresponding survival probability of $p(x, t \mid \delta)$ in this 2AFC scenario is:

$$S^*(t \mid \delta, \gamma_u, \gamma_l) = \int_{\gamma_l}^{\gamma_u} p(x, t \mid \delta) dx = \int_{-\infty}^{\gamma_u} p(x, t \mid \delta) dx - \int_{-\infty}^{\gamma_l} p(x, t \mid \delta) dx$$

$$= S(t \mid \delta, \gamma_u) - S(t \mid \delta, \gamma_l)$$

This algebraic decomposition highlights that $S^*(t \mid \delta, \gamma_u, \gamma_l)$ will necessarily transition into $S(t \mid \delta, \gamma_u)$ if $S(t \mid \delta, \gamma_l)$ becomes negligibly small because of an extremely low chance that the process hits $\gamma_l$. Thus, a Wiener distribution of response times can be well approximated by the Wald distribution, whenever these data predominantly arise from one out of two available response options.

**Online Supplement**

The presented raw data of the Go/No-go task (6 participants) and the R syntax to fit the censored shifted Wald and the Wiener distribution are provided online as supplementary file (*CensWald_GoNogo.zip*) to this article.

**Acknowledgements**

This work was partly supported by the German Research Foundation (SFB 940/1). We thank Andrew Heathcote, Dominik Wabersich, Dóra Erbé-Matzke, Matthias Gondan, and Monika Fleischhauer for their most valuable support and comments on the manuscript.

**References**

Anders, R., Alario, F.-X., & Van Maanen, L. (2016). The shifted Wald distribution for response time analysis. *Psychological Methods*, 21, 309-327.

Balota, D. A., Yap, M. J. (2011). Moving beyond the mean in studies of mentral chronometry: the power of response time distributional analyses. *Current Directions in Psychological Science*, 20, 160-166.

Brown, S. D., Heathcote, A. (2008). The simplest complete model of choice response time: Linear ballistic accumulation. *Cognitive Psychology*, 57, 153-178.

Cleveland, W. S., Devlin, S. J. (1988). Locally weighted regression: an approach to regression analysis by local fitting. *Journal of the American Statistical Association*, 83, 596-610.

Gomez, P., Ratcliff, R., & Perea, M. (2007). A model of the go/no-go task. *Journal of Experimental Psychology: General*, 136, 389-413.

Gondan, M., Blurton, S. P., & Kesselmeier, M. (2014). Even faster and even more accurate first-passage time densities and distributions for the Wiener diffusion model. *Journal of Mathematical Psychology*, 60, 20-22.

Heathcote, A. (2004). Fitting Wald and ex-Wald distributions to response time data: an example using functions for the S-PLUS package. *Behavior Research Methods, Instruments, & Computers*, 36, 678-694.

Jones, M., & Dzhafarov, E. N. (2014). Unfalsifiability and mutual translatability of major modeling schemes for choice reaction time. *Psychological Review*, 121, 1-32.

Laming, D. R. J. (1968). *Information theory of choice-reaction times*. Academic Press: Oxford.

Link, S. W., & Heath, R. A. (1975). A sequential theory of psychological discrimination. *Psychometrika*, 40, 77-105.

Logan, G. D., Van Zandt, T., Verbruggen, F., & Wagenmakers, E. J. (2014). On the ability to inhibit thought and action: General and special theories of an act of control. *Psychological Review*, 121, 66-95.

Lord, F. M. (1953). The relation of test score to the trait underlying the test. *Educational and Psychological Measurement*, 13, 517-548.

Matzke, D., & Wagenmakers, E.-J. (2009). Psychological interpretation of the ex-Gaussian and shifted Wald parameters: a diffusion model analysis. *Psychonomic Bulletin & Review*, 16, 798-817.

Miller, R. G. (1998). *Survival analysis*. 2$^{nd}$ ed. John Wiley & Sons: New York.

Navarro, D. J., Fuss, I. G. (2009). Fast and accurate calculations for first-passage times in Wiener diffusion models. *Journal of Mathematical Psychology*, 53, 222-230.

Nelder, J. A., & Mead, R. (1965). A simplex algorithm for function minimization. *Computer Journal*, 7, 308–313.

Prentice, R. L., Kalbfleisch, J. D., Peterson, A. V., Flournoy, N., Farewell, V. T., & Breslow, N. E. (1978). The analysis of failure times in the presence of competing risks. *Biometrics*, 34, 541-554.

Pratte, M. S., Rouder, J. N., Morey, R. D., & Feng, C. (2010). Exploring the differences in distributional properties between Stoop and Simon effects using delta plots. *Attention, Perception, & Psychophysics*, 72, 2013-2025.

R Core Team (2014). *R: A language and environment for statistical computing*. R Foundation for Statistical Computing: Vienna.

Ratcliff, R. (1978). A theory of memory retrieval. *Psychological Review*, 85, 59-108.

Ratcliff, R., & Rouder, J. N. (1998). Modeling response times for two-choice reaction times. *Psychological Science*, 9, 347-356.

Ratcliff, R., & Tuerlinckx, F. (2002). Estimating parameters of the diffusion model: approaches to dealing with contaminant reaction times and parameter variability. *Psychonomic Bulletin & Review,* 9, 438-481.

Siannis, F., Copas, J., & Lu, G. (2005). Sensitivity analysis for informative censoring in parametric survival models. *Biostatistics*, 6, 77-91.

Stone, M. (1960). Models for choice-reaction time. *Psychometrika*, 25, 251-260.

Trueblood, J. S., & Endres, M. J., Busemeyer, J.R., & Finn, P. R. (2011). Modeling response times in a go/no-go discrimination task. *CogSci 2011 Proceedings*, 1866-1871. Retrieved from https://mindmodeling.org/cogsci2011/papers/0416/paper0416.pdf

Ulrich, R., & Miller, J. (1994). Effects of truncation on reaction time analysis. *Journal of Experimental Psychology: General*, 123, 34-80.

Vandekerckhove, J., Tuerlinckx, F., & Lee, M. D. (2011). Hierarchical diffusion models for two-choice response times. *Psychological Methods*, 16, 44-62.

Verbruggen, F., & Logan, G. D. (2008). Automatic and controlled response inhibition: associative learning in the go/no-go and stop-signal paradigms. *Journal of Experimental Psychology: General*, 137, 649-672.

Van Ravenzwaaij, D., & Oberauer, K. (2009). How to use the diffusion model: Parameter recovery of three methods: EZ, fast-dm, and DMAT. *Journal of Mathematical Psychology*, 53, 463-473.

Van Zandt, T., & Ratcliff, R. (1995). Statistical mimicking of reaction time data: Single-process models, parameter variability, and mixtures. *Psychonomic Bulletin & Review*, *2*, 20–54.

Van Zandt, T. (2000). How to fit a response time distribution. *Psychonomic Bulletin & Review*, 7, 424-465.

Voss, A., Nagler, M., & Lerche, V. (2013). Diffusion models in experimental psychology: a practical introduction. *Experimental Psychology*, 60, 385-402.

Wabersich, D., & Vandekerckhove, J. (2014). The RWiener package: an R package providing distribution functions for the Wiener diffusion model. *The R Journal*, 6, 49-56.

Wagenmakers, E.-J., van der Maas, H. L. J., & Grasman, R. P. P. P. (2007). An EZ-diffusion model for response time and accuracy. *Psychonomic Bulletin & Review*, 14, 3-22.

Wagenmakers, E.-J., van der Maas, H. L. J., Dolan, C. V., & Grasman, R. P. P. P. (2008). EZ does it! Extensions of the EZ-diffusion model. *Psychonomic Bulletin & Review*, 15, 1229-1235.

Wagenmakers, E.-J. (2009). Methodological and empirical developments for the Ratcliff diffusion model of response times and accuracy. *European Journal of Cognitive Psychology*, 21, 641-671.

Table 1. Illustration of the violation of the specific objectivity of parameter comparisons between different individuals due to inappropriate constraints of the diffusion model. Each of three fictive individuals, who differ in their dispositional response cautions ($\alpha$) and drift rates ($\delta$), completes two conditions (i.e., sets of identical items / trials) of a 2AFC task. The experimental manipulation that determines these two conditions selectively alters the response tendency ($\beta$). The psychometric objective is the recovery of the between-individual/condition differences in the data-generating (i.e., "true") diffusion parameters configurations by fitting the Wiener distribution.

|  | Condition | True diffusion parameters | | | | Performance characteristics | | | Recovered diffusion parameters | | | |
|---|---|---|---|---|---|---|---|---|---|---|---|---|
|  |  | $\alpha$ | $\beta$ | $\delta$ | $\theta$ | M(RT) | SD(RT) | %Correct | $\alpha$ | $\beta$ | $\delta$ | $\theta$ |
| Bob | A | 2 | 0.5 | 3 | 0.3 | 0.63 s | 0.19 s | 99.7 | 2 | 0.5 | 3 | 0.3 |
|  | B | 2 | 0.8 | 3 | 0.3 | 0.43 s | 0.12 s | > 99.9 | 1.2 | 0.5 | 3.8 | 0.3 |
| Stephen | A | 2.5 | 0.5 | 3 | 0.3 | 0.71 s | 0.21 s | > 99.9 | 2.5 | 0.5 | 3 | 0.3 |
|  | B | 2.5 | 0.8 | 3 | 0.3 | 0.47 s | 0.14 s | > 99.9 | 1.3 | 0.5 | 3.5 | 0.3 |
| Dan | A | 2 | 0.5 | 4 | 0.3 | 0.55 s | 0.13 s | > 99.9 | 2 | 0.5 | 4 | 0.3 |
|  | B | 2 | 0.8 | 4 | 0.3 | 0.40 s | 0.08 s | > 99.9 | 1 | 0.5 | 4.7 | 0.3 |

*Note.* Due to the overall very low error probabilities, $\beta$ was constrained to 0.5. Accordingly, the individual differences between the diffusion parameters of Bob, Stephen, and Dan could be fully recovered in condition A. By contrast, the recovered response caution in condition B differed between Bob and Dan ($\alpha_{Bob} > \alpha_{Dan}$), although it was actually identical. Similarly, the drift rate in condition B differed between Bob and Stephen ($\delta_{Bob} > \delta_{Stephen}$), although it was actually identical.

Table 2. The relation between shifted Wald and Wiener parameters, and their psychological interpretations.

| Shifted Wald parameter | Wiener parameter configuration | Psychological interpretation | Simulation lower bound | Simulation upper bound |
|---|---|---|---|---|
| $\delta$ | $\delta$ | Speed of evidence accumulation for the correct response option | $0.00$ s$^{-1}$ | $5.86$ s$^{-1}$ |
| $\gamma$ | $\alpha(1-\beta)$ | Speed-accuracy tradeoff relative to the starting point of evidence accumulation | $0.28$ | $1.97$ |
| $\theta$ | $\theta$ | Time that is not related to evidence accumulation (e.g., sensory encoding or motor processes) | $0.21$ s | $0.94$ s |

*Note.* As outlined in section 2.1., the boundaries of the Wiener distribution are commonly defined by $\alpha$ (upper boundary) and 0 (lower boundary), with the starting point of the diffusion process at $\beta$ (i.e., the relative initial response tendency towards the correct response option). Translating this parameterization, the distances to the boundaries for the correct and erroneous response are defined as $\gamma_u = \alpha(1-\beta)$ and $\gamma_l = \alpha\beta$, respectively (see also Van Zandt, 2000). The parameter ranges reported in the last two columns have been accordingly converted from Matzke and Wagenmakers (2008) assuming $\beta = 0.5$.

Table 3. Performance data and diffusion parameters as estimated from the Go-trial RTs of six participants by conventional and censored shifted Wald modeling.

| ID | Descriptive statistics | | | | | | | sW parameters | | | csW parameters | | | LR |
|----|----|----|----|----|----|----|----|----|----|----|----|----|----|----|
|    | M  | SD | %C | Min | Q50 | Max | | δ | γ | θ | δ | γ | θ | |
| 1 | 319 | 53.6 | 100 | 211 | 309 | 674 | | 7.92 | 1.19 | 0.17 | 7.92 | 1.19 | 0.17 | 1 |
| 2 | 350 | 62.8 | 100 | 235 | 337 | 634 | | 6.86 | 1.24 | 0.17 | 6.86 | 1.24 | 0.17 | 1 |
| 3 | 343 | 63.2 | 100 | 222 | 333 | 731 | | 7.11 | 1.27 | 0.16 | 7.11 | 1.27 | 0.16 | 1 |
| 4 | 399 | 91.0 | 100 | 254 | 374 | 837 | | 4.92 | 0.88 | 0.22 | 4.92 | 0.88 | 0.22 | 1 |
| 5 | 354 | 99.2 | 100 | 202 | 327 | 869 | | 4.64 | 0.89 | 0.16 | 4.64 | 0.89 | 0.16 | 1 |
| 6 | 299 | 44.1 | 100 | 215 | 295 | 610 | | 9.00 | 1.30 | 0.15 | 9.00 | 1.30 | 0.15 | 1 |
| 1 | 316 | 45.6 | 98.9 | 211 | 308 | 479 | | 9.25 | 1.52 | 0.15 | 8.23 | 1.26 | 0.17 | 0.036 |
| 2 | 346 | 54.6 | 97.7 | 235 | 335 | 498 | | 8.41 | 1.79 | 0.13 | 6.98 | 1.28 | 0.17 | 0.018 |
| 3 | 338 | 52.2 | 97.7 | 222 | 330 | 498 | | 8.97 | 1.91 | 0.13 | 7.45 | 1.38 | 0.16 | 0.002 |
| 4 | 374 | 51.1 | 88.3 | 254 | 368 | 496 | | 9.69 | 2.42 | 0.12 | 5.45 | 1.00 | 0.21 | <0.001 |
| 5 | 330 | 61.9 | 90.6 | 202 | 322 | 498 | | 7.70 | 1.75 | 0.10 | 4.69 | 0.90 | 0.16 | <0.001 |
| 6 | 298 | 40.8 | 99.7 | 215 | 295 | 497 | | 9.97 | 1.61 | 0.14 | 9.35 | 1.41 | 0.15 | 0.514 |

*Note.* The lower part of the table has been derived by truncating the original RT data (grey cells) at 500 ms. Thus artificial omission errors were generated. %C = percent correct RTs. RTs are scaled in ms. Diffusion parameters were estimated from RTs scaled in seconds. sW = conventional shifted Wald, csW = censored shifted Wald, LR = likelihood ratio of the sW and the csW parameters given the original RTs.